# Impact of solar activity on climate changes in Athens region, Greece

Nectaria A. B. GIZANI*, Kimon PAPATHANASOPOYLOS, Leonidas VATIKIOTIS, Efthimios ZERVAS

Tel: +30-2610-367-521
Fax : +30-2610-367-520
e-mail: ngizani@eap.gr

Address
Hellenic Open University, School of Sciences and Technology
Tsamadou 13-15 & Ag. Andreou, 26222
Patra, Greece

**Abstract:** The scope of this work is to study the role that the solar weather plays in terrestrial weather. For this reason we study the effect of the solar activity on the climate changes in Greece. In the current work we look for possible correlation between the solar activity data spanning the years from 1975 to 2000 and the meteorological data from two weather stations based inside the city of Athens, Greece (New Philadelphia) and in greater Athens in the north of Attica (Tatoi area). We examine the annual variations of the average values of six meteorological parameters: temperature, atmospheric pressure, direction and intensity of wind, rainfall and relative air humidity. The solar data include decade variations, within the above period, of the solar irradiance, mean sunspot number between two solar cycles, magnetic cycle influence, and solar UV driving of climate (radio flux).

**Keywords**: Climate changes, indicators, solar activity

## 1. Introduction

The Sun, the only star of our planetary system, is the main source of energy supply for Earth. The Sun is a unique laboratory of magnetized plasmas. Intense outflows of ionized particles, mostly protons and electrons in the interplanetary space, form the solar wind. An increase in solar activity (for a simplified picture this implies more sunspots) is accompanied by an increase in the solar wind. In turn the solar wind forms the heliosphere in which Earth and the rest of our planets bathe. A more active solar wind and stronger magnetic field reduces the cosmic rays striking the Earth's atmosphere. The ionizing solar particles enter the magnetosphere, ionosphere and upper atmosphere, producing auroras, magnetic-storms and other phenomena known as space weather. These changes in the solar radiative output have an impact on the terrestrial energy balance and can affect the chemistry of the stratosphere.

Observations and theory indicate that the Sun is variable on all time scales, from seconds to centuries depending on the time-scale of interest. The most common cycle of the solar variability is the ≈11.1 yr cycle (the average value of 9–13 years), or sunspot cycle. This is the periodicity between successive manifestations of the solar activity. When the inversion of polarization of the polar magnetic field on the Sun is considered the period of variability has a 22-year period. That is because the magnetic field of the Sun reverses during each sunspot cycle, so the magnetic poles have the same charge (positive/negative) after two reversals. There are longer periods than these spanning 80–100 years.

All solar phenomena, relevant to the terrestrial climate (e.g. solar irradiance variations, coronal mass ejections etc), are magnetically driven. Therefore a better understanding of the Sun-Earth system requires a good knowledge of the Sun's magnetic field from the solar core to the corona and heliosphere, as well as the magnetic field's evolution on timescales from minutes to millennia.

The Sun's magnetic variability is thought to contribute to the global climate change through three mechanisms (Figure 1): (i) Variations in the total solar irradiance (TSI, integrated over all wavelengths ) can cause changes in the energy input into the Earth's atmosphere. We mention here that the near-UV, visible, and near-IR irradiance can directly affect the Earth's radiative balance and surface temperature. They constitute ≈ 99% of the Sun's total radiative output and penetrate the terrestrial atmosphere to the troposphere, reaching the surface; (ii) Variations of the solar UV irradiance, can alter chemical and physical processes in the Earth's upper atmosphere (stratospheric chemistry) . Changes in the stratosphere provide indirect climate effect, since the latter is coupled to the troposphere; (iii) Changes in low cloud cover can be induced by the solar cosmic ray flux reaching Earth, modulated by cyclic variations of the Sun's open magnetic flux (e.g. Marsh and Svensmark, 2000).

Solar irradiance is expected to follow the 11-year solar activity cycle. Krivova et al. 2009 for example have found that it varies by about 0.1% during the cycle. Almost 31% of the incident solar radiation is reflected back into space, 20% of which is absorbed by the lower atmosphere, and 49% is absorbed by the surface and oceans (Kiehl and Trenberth, 1997).

TSI measurements exist since 1978. They are in a 'patch-form' since they were taken with different instruments each with its own calibration, resulting in different scientific conclusions. The sunspot number is used as a proxy for the solar irradiance (e.g. Lean et al., 1995).

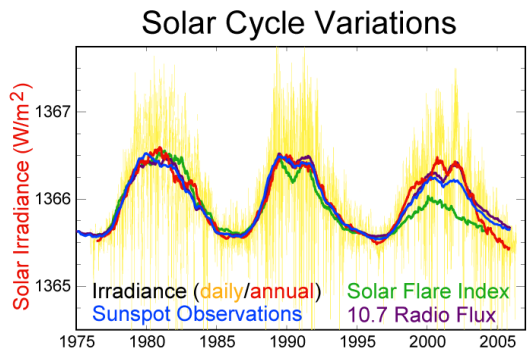

Figure 1. Monthly averages of different proxies of the solar magnetic activity. The color code describes the variables: Irradiance, sunspot number, radio flux at 10.7 cm indicating the UV-irradiance, solar activity in the form of flares. Data compiled from data taken from literature (e.g. Solanki and Krivova, 2003).

About 15 Wm$^{-2}$ (~1%) of the solar UV-radiation is absorbed by the terrestrial atmosphere. UV-radiation does not affect directly the climate as it is not supposed to reach the Earth's surface. Its effect is through the coupling of the stratosphere with the troposphere: UV radiation photo-dissociates molecular oxygen creating the ozone layer (Haigh et al. 2004). Ozone resides primarily in the stratosphere controlling the deposition of the solar energy in the stratosphere. Variations in the ozone amount alter the altitudinal temperature gradient from the troposphere to the stratosphere, and the latitudinal temperature gradient in the stratosphere from the equator to the poles.

UV irradiance changes by up to 10% in the 150–300 nm range and by an amount greater than 50% at shorter wavelengths, including the Ly-α emission line near 121.6 nm (e.g. Krivova et al. 2009 and references therein). It is more variable than TSI by an order of magnitude or so. It contributes significantly to the changes in TSI (15% of the TSI cycle, Lean et al., 1997). Krivova and Solanki 2004 find that the average 11 year UV irradiance has a very similar form to the TSI between 1856 and 1999. However the relative change is greater for the UV irradiance.

There is a linear correlation between the annual mean irradiance and the radio flux of the Sun at 10.7 cm. Solar flares or coronal mass ejections, can trigger shock waves which leave signatures in radio spectrograms in the metric range (corona) up to the kilometric range (interplanetary medium) as they propagate in the solar corona or in the interplanetary medium. The F10.7 index is used as a proxy for the UV and EUV spectral irradiance, which varies day by day since it depends on the number of active regions on the Sun. This proxy shows the solar influence on terrestrial parameters such as the temperature at a given atmospheric layer. Radio data are taken daily since 1947. Mg II core-to-wing ratio obtained on-board of several spacecraft is also used as a proxy of UV irradiance (e.g. Viereck and Puga, 1999).

The third mechanism is affecting the terrestrial climate indirectly as well. Galactic cosmic ray (GCR) flux is suggested to affect tropospheric temperature via cloud-cover variations (low clouds have a strong cooling effect on climate). The physical mechanism is the following: GCR particles are essentially the source of ionization in the troposphere above 1 km. Changes in ionization affect the abundance of aerosols that serve as the nuclei of condensation for cloud formation. GCR flux varies inversely with solar activity. Reliable estimate of the evolution of the Sun's magnetic field, modulating cosmic rays, exist since 1868 (e.g. Solanki et al., 2000). This solar variable is measured directly with sufficient quality by Neutron Monitors since 1953. The data missing are reconstructed by converting the open magnetic flux and neutron monitor data into cosmic-ray flux. Krivova and Solanki, 2004 have found that the two quantities follow each other closely up to 1985, and diverge strongly after that.

Data interpretation indicates significant contribution of the solar variability to climate change is significant and that the temperature trend since 1980 can be large and upward on a time-scale of decades to centuries (e.g. Solanki and Fligge, 1999). Reconstruction models such as PCMOD and ACRIM (e.g. Lockwood 2008) estimate that the solar contribution to global warming is negligible since 1980. The Intergovernmental Panel for Climate Change (IPCC) 2007 claims that the solar contribution to climate change is negligible since 1950. However Scafetta (2009) suggested that IPCC 2007 could be using wrong solar data as well as models which underestimate several climate mechanisms. Taking into account the entire range of possible total solar irradiance (TSI) satellite composite since 1980, the author finds that the solar contribution to climate change ranges from a slight cooling to a significant warming, which can be as large as 65% of the total observed global warming.

Global temperature fall slightly after 1950 until the mid-1970s, and has risen sharply since, and particularly after 1985. This warming has been attributed to human activity by several authors (e.g., Mitchell et al., 2001). Solar irradiance seems to have decreased slightly in this period. A correlation between the cosmic-ray intensity and the coverage of low-lying clouds over a solar cycle has been suggested by Marsh and Svensmark (2000). Correlation between variations in solar irradiance and low cloud cover (affected through the atmospheric circulation) seems to be stronger than the correlation between galactic cosmic rays and low clouds. Kristjánsson et al., (2002) and (2004) demonstrated a large negative correlation between total solar irradiance and low cloud cover for the whole period 1983–1999, with no discrepancies after 1994. This implies that any positive correlation between GCR flux and low cloud cover would be coincidental. They argue that it is of no significance for the present global warming anyway, since solar irradiance appears to be almost constant over the last 50 years. Lockwood et al., (1999) assumed that GCR flux has decreased significantly over the 20th century (i.e. the solar magnetic field increased) although the results were challenged by Richardson et al. (2002), who found no significant trends for either sunspots, GCR flux or the interplanetary magnetic field over the past 50 years. However positive trends in sunspot number and magnetic field do seem to exist. Marsh and Svensmark, (2000) suggested that GCR flux increase resulted in the decrease of low cloud cover by more than 8%, accounting for a significant fraction of the observed global warming.

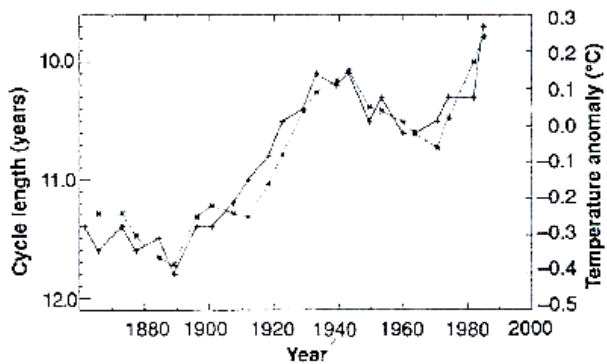

Figure 2. Land temperature of Northern Hemisphere (stars) and solar cycle length (inverted, pluses).
Both time series are smoothed with a (1 2 2 2 1) filter weightings. Reproduction from Friis-Christensen and Lassen (1991) by Haigh, 2007.

To sum up the role that the Sun plays on the climate is being a subject of intense debate. Reconstruction models are based also on stellar observations, attempt to describe the past solar behavior on the terrestrial climate with contradicting results. Solar data suffer from bad quality in the past years (uncertainties in historical data of TSI for example) and of different instrument calibration in the satellite era. However there is a current need to distinguish between natural and anthropogenic causes of climate change. Therefore the study of the effect of the solar variability on the terrestrial climate is to the forefront of meteorological research. There is statistical evidence for solar influence on various meteorological parameters on all timescales (eg. see Fig. 2). However extracting the signal from the noise in a naturally highly variable system remains a key problem.

In the current research work we present preliminary results from the effect of the parameters of the solar activity on climate in Attica, Greece. We use several meteorological parameters from two weather stations based inside the city of Athens, Greece (New Philadelphia) and in the greater Athens in the north of Attica (Tatoi area). The solar variables studied are the TSI, UV-solar irradiance, cosmic ray flux and sunspot number, averaged over the 11-year solar cycle spanning the years from 1975 to 2000. The meteorological parameters are the temperature, atmospheric pressure, direction and force of wind, relative air humidity and precipitation. We use several statistical parameters as indicators of the climate change: the annually averaged value of each meteorological parameter; the error on the mean (standard deviation); the skewness; the kurtosis of the annual distribution of each of the above six meteorological parameters.

## 2. Methodology

Solar data span two 11- year solar cycles from 1975 up to 2005. The plot of all solar proxies is added for comparison at the end of every meteorological parameter graph. It is adapted from Solanki et al., 2006. The solar variables plotted are from top to bottom the TSI, F10.7 cm index as an indicator of the UV-irradiance, the magnitude solar flux, and the mean sunspot number. Not plotted here is the GCR flux, for which we remind the reader that this flux demonstrates an almost inversed behavior to the solar activity.

All meteorological data used in this work come from the Hellenic National Meteorological Service (EMY). Those data come from two stations on the Attiki area, one in Athens city (Nea Filadelfeia) and one in the northern of Athens (Tatoi area). The data cover 45 years from 1956 to 2001. For this work, as the solar data cover the period after 1975, the data covering the years 1956-1975 are ignored. Each station measures 6 meteorological parameters: temperature (°C), atmospheric pressure (mm Hg), wind direction (°, with 0° to the north), wind force (knot), relative humidity (%) and precipitation (mm). For all data a point is measured every 3 hours (starting from midnight) except precipitation which is a 12 hours average. For all meteorological parameters the following statistical parameters are calculated for all years: average value, standard deviation, skewness and kurtosis (see Savvidou et al., these proceedings, for an explanation on these parameters).

## 3. Results and Discussion

### 3.1 Impact of solar activity on average annual values of meteorological parameters

Figures 3-8 show from bottom to top the evolution of four solar parameters: mean sunspot number, magnetic flux, radio flux, and solar irradiance over time for the years 1975-2003. The four solar parameters show a cyclic variability with 3 maxima at about 1980, 1991 and 2002 and three minima at about 1975, 1986 and 1997.

Our simple analysis, although preliminary, does not suggest a correlation between the four solar parameters and the average annual values of the six meteorological parameters of the two stations in Athens area.

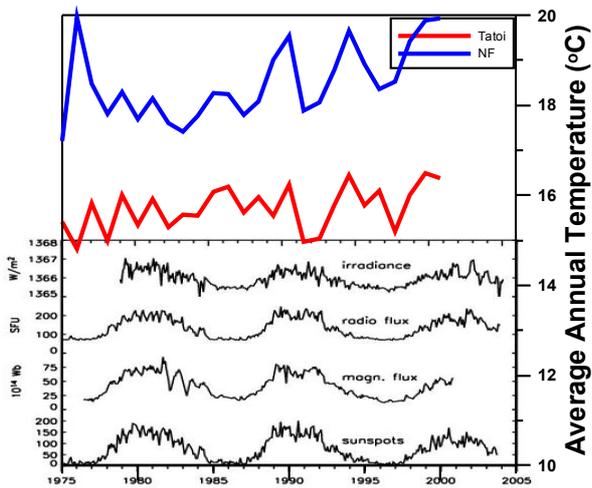

Figure 3. Average annual values of temperature for the two stations in Athens area and values of the four solar parameters studied here.

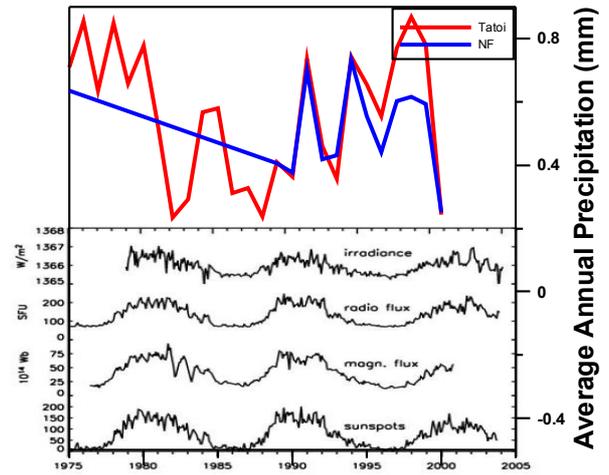

Figure 6. Average annual values of precipitation for the two stations in Athens area and values of the four solar parameters studied here.

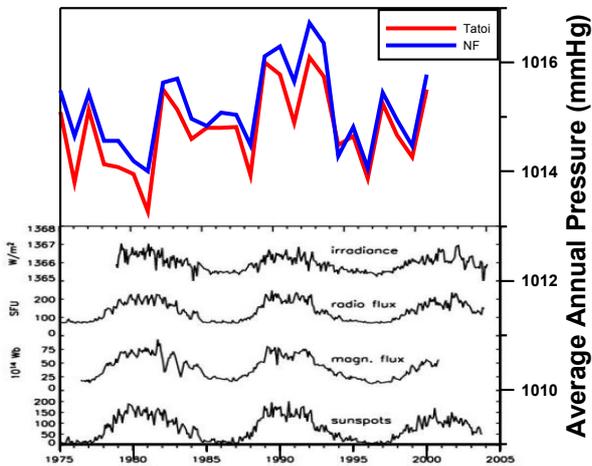

Figure 4. Average annual values of atmospheric pressure for the two stiosn in Athens area and of the four solar parameters studied here.

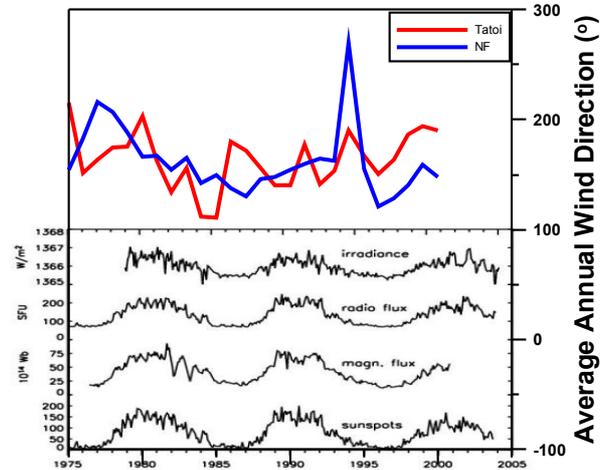

Figure 7. Average annual values of wind direction for the two stations in Athens area and of the four solar parameters studied here.

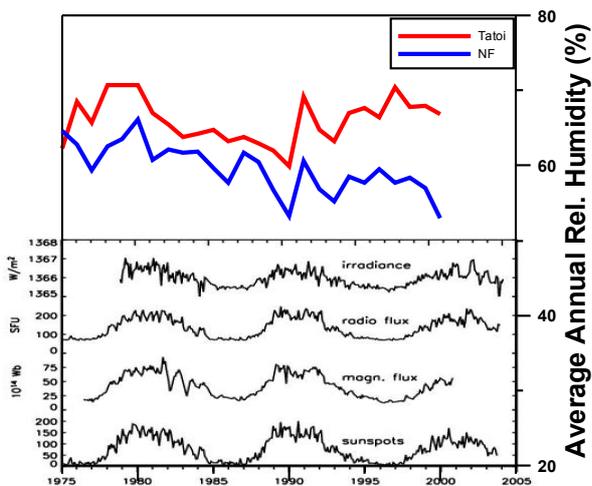

Figure 5. Average annual values of relative humidity for the two stations in Athens area and of the four solar parameters studied here.

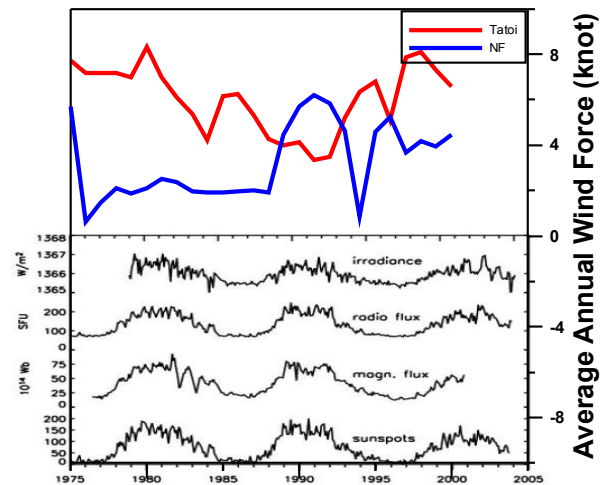

Figure 8. Average annual values of wind force for the two stations in Athens area and values of the four solar parameters studied here.

### 3.2 Impact of solar activity on annual standard deviation, skewness and kurtosis of meteorological parameters

Current bibliography studies the impact of solar activity on global climate and data suggest a potential influence (eg. Haigh 2007). However results have been controversial. In the previous section we found that there is no obvious correlation between the four solar parameters and the average value of the six meteorological parameters in Athens area. The next step is to look if there is a correlation between the four solar parameters and the statistical characteristics of the six meteorological parameters. Figures 9-14 show the evolution of the four solar parameters in time and the standard deviation of the annual values of the six meteorological parameters; figures 15-20 the evolution of the same parameters and the average of annual data of the six parameters; figures 21-26 the same correlations using the kurtosis of the same data.

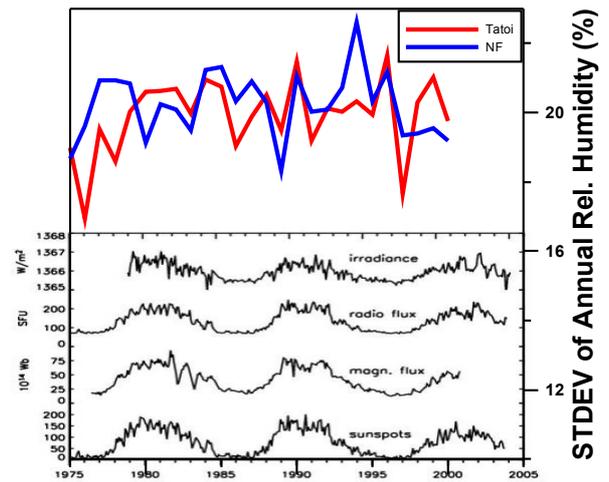

Figure 11. Average annual standard deviation of relative humidity for the two stations in Athens area and of the four solar parameters studied here.

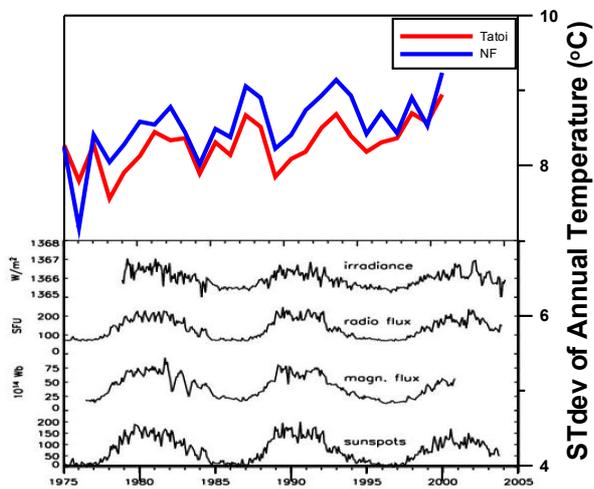

Figure 9. Average annual standard deviation of temperature for the two stations in Athens area and of the four solar parameters studied here.

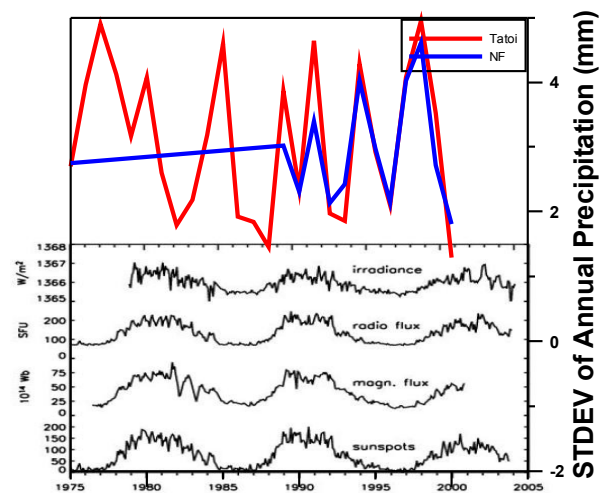

Figure 12. Average annual standard deviation of precipitation for the two stations in Athens area and values of the four solar parameters studied here.

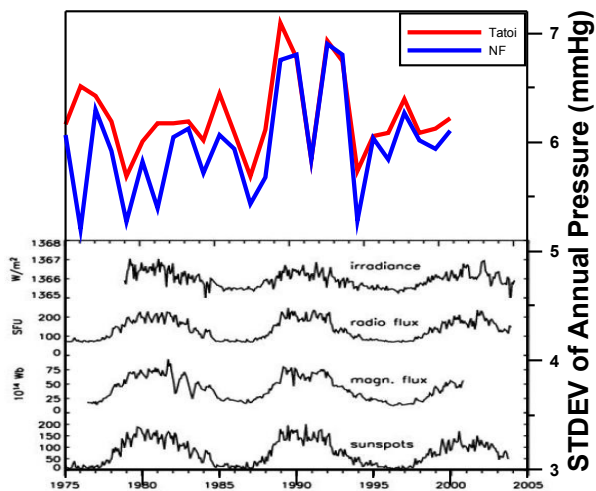

Figure 10. Average annual standard deviation of atmospheric pressure for the two stations in Athens area and values of the four solar parameters studied here.

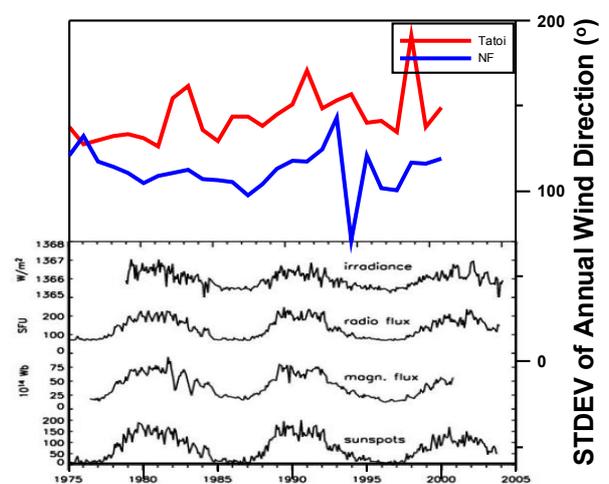

Figure 13. Average annual standard deviation of wind direction for the two stations in Athens area and of the four solar parameters studied here.

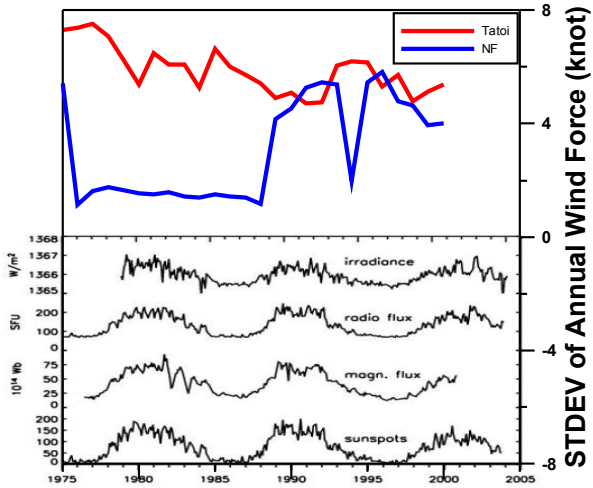

Figure 14. Average annual standard deviation of wind force for the two stations in Athens area and values of the four solar parameters studied here.

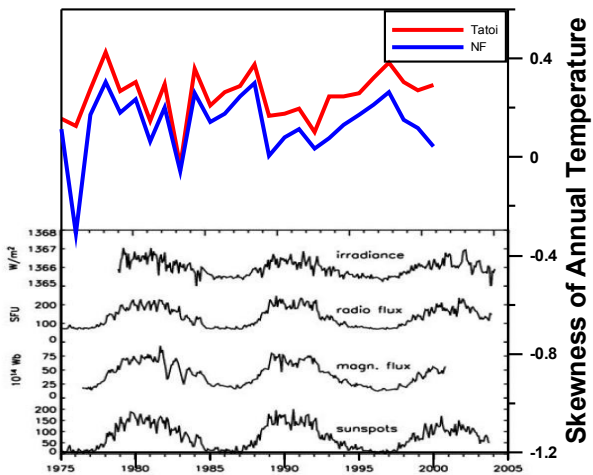

Figure 15. Annual skewness of temperature or the two stations in Athens area and of the four solar parameters studied here.

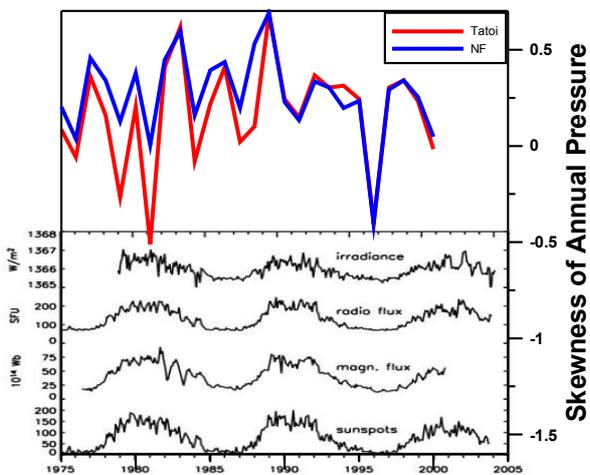

Figure 16. Annual skewness of atmospheric pressure for the two stations in Athens area and values of the four solar parameters studied here.

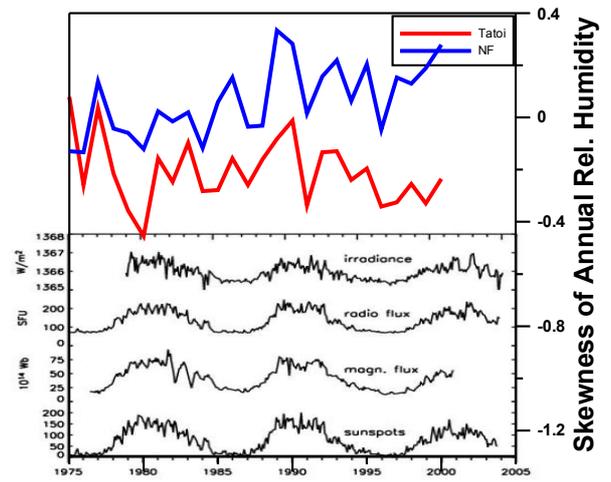

Figure 17. Annual skewness of relative humidity for the two stations in Athens area and of the four solar parameters studied here.

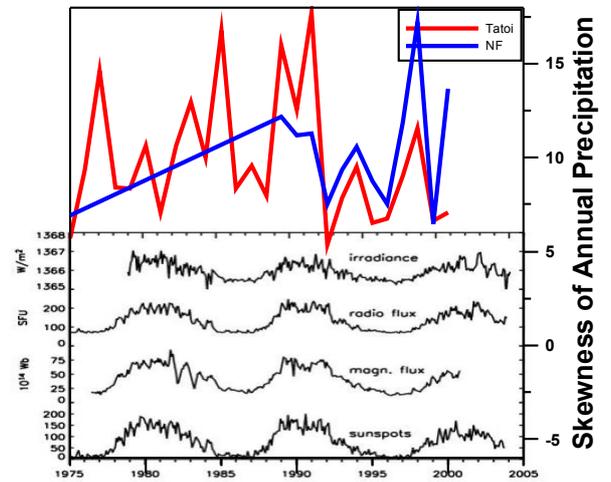

Figure 18. Annual skewness of precipitation for the two stations in Athens area and values of the four solar parameters studied here.

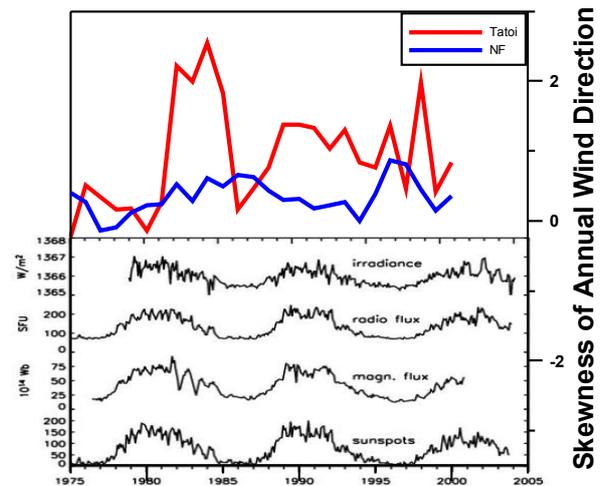

Figure 19. Annual skewness of wind direction for the two stations in Athens area and of the four solar parameters studied here.

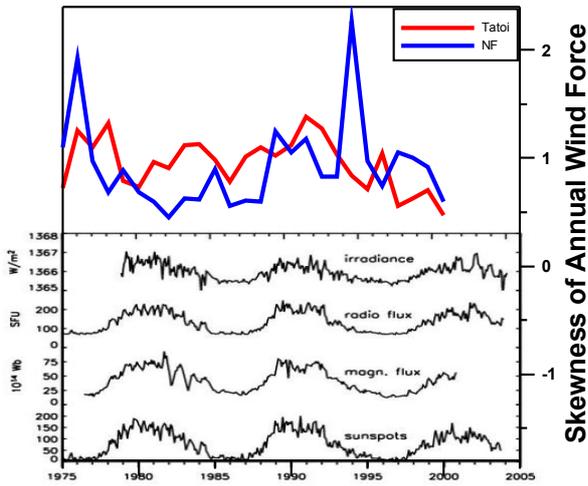

Figure 20. Annual skewness of wind force for the two stations in Athens area and of the four solar parameters studied here.

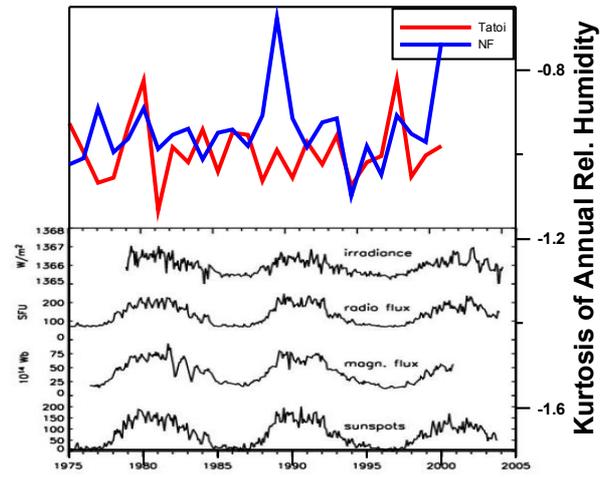

Figure 23. Annual kurtosis of relative humidity for the two stations in Athens area and of the four solar parameters studied here.

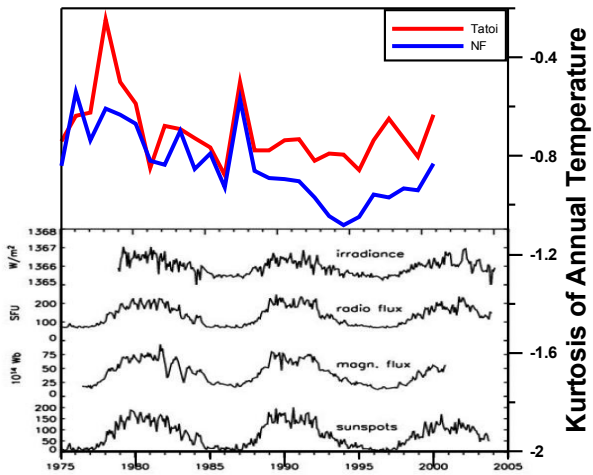

Figure 21. Annual kurtosis of temperature for the two stations in Athens area and of the four solar parameters studied here.

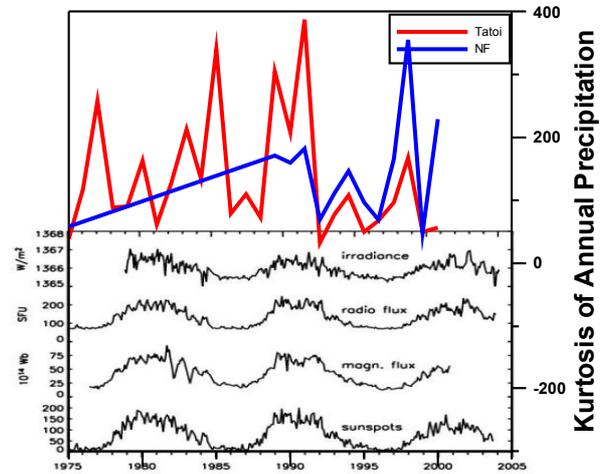

Figure 24. Annual kurtosis of precipitation for the two stations in Athens area and values of the four solar parameters studied here.

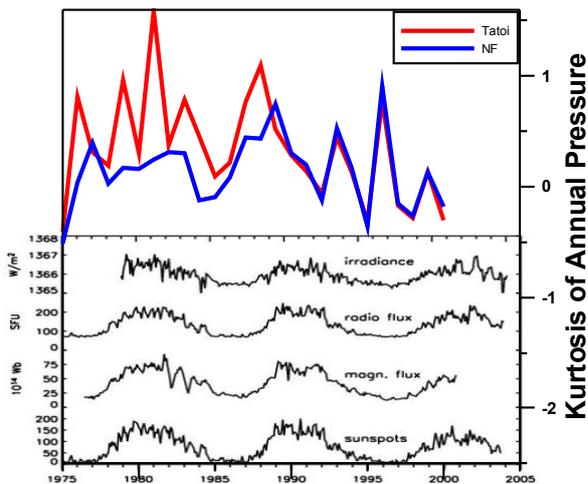

Figure 22. Annual kurtosis of atmospheric pressure for the two stations in Athens area and values of the four solar parameters studied here.

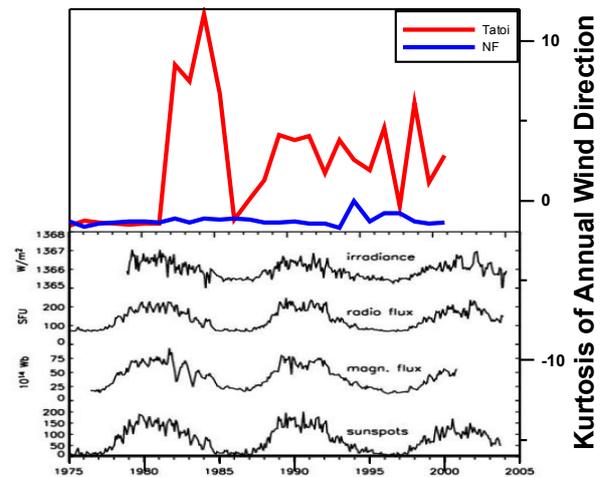

Figure 25. Annual kurtosis of wind direction for the two stations in Athens area and of the four solar parameters studied here.

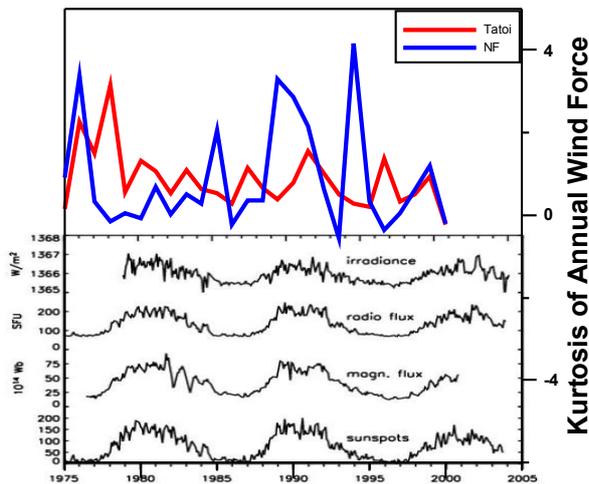

Figure 26. Annual kurtosis of wind force for the two stations in Athens area and of the four solar parameters studied here.

Similarly, as in the case of annual average values and solar proxies of solar activity are not correlated in an obvious manner. However, this is only a partial study, focused only in the greater Athens area. The correlations are probably altered due to urban and semi-urban conditions of the examined area.

Next step is to do a more sophisticated analysis which will include extracting the data from noise, use appropriate smoothing and simulating average temperature changes in Greece in response to climate forcings with for example an 1-D Energy Balance Model (EBM). We will also use more meteorological stations scattered all over Greece and the solar activity proxies TSI, UV-irradiance, F10.7 index, sunspot number, cosmic-ray flux/low cloud cover.

Another characteristic of meteorological data is the evolution of extreme values. The possible correlation of that characteristic with solar activity will also be examined.

## 3. Conclusions

In the current paper we present preliminary results of the effect that the solar variables-indicators of the solar activity have on climate characteristics in Attica, Greece.

The solar variables used were TSI, UV-solar irradiance, cosmic ray flux and sunspot number, averaged over two 11-year solar cycle spanning the years from 1975 to 2000. Meteorological parameters include temperature, atmospheric pressure, direction and force of wind, relative air humidity and precipitation data taken from two weather stations, one inside the city (Nea Filadelfeia) and one some kilometers outside the city, in the northern region of Attica (Tatoi).

Although Sun has the potential to affect climate we have found no particular trend between each set of parameters. However we suggest that this result is probably due to urban conditions. Our future work will include data from other stations in Greece (from the National Meteorological Service) from the north to the south (islands) together with analyzed solar variables (TSI, UV-irradiance, F10.7 index, sunspot number, cosmic-ray flux/low cloud cover).